\documentclass[preprint,sort&compress,5p,twocolumn]{elsarticle}
\usepackage{CJK}
\usepackage{amsmath,amssymb}
\usepackage{mathtools,extarrows}
\usepackage{float,subfigure,tikz,tikzscale,hyperref}
\usepackage{rotating}

\journal{Physics Letters B}
\usetikzlibrary{calc}
\usetikzlibrary{decorations,snakes}

\begin{document}

\title{A universal splitting of tree-level string and particle scattering amplitudes}

\author[1,2]{Qu Cao}
\ead{qucao@zju.edu.cn}
\author[1,3]{Jin Dong}
\ead{dongjin@itp.ac.cn}
\author[1,4,5]{Song He}
\ead{songhe@itp.ac.cn}
\author[1]{Canxin Shi}
\ead{shicanxin@itp.ac.cn}

\affiliation[1]{organization={CAS Key Laboratory of Theoretical Physics,
    Institute of Theoretical Physics, Chinese Academy of Sciences},
    city={Beijing},
    postcode={100190},
    country={China}}
\affiliation[2]{organization={Zhejiang Institute of Modern Physics, Department of Physics, Zhejiang University}, 
    city={Hangzhou},
    postcode={310027},
    country={China}}
\affiliation[3]{organization={School of Physical Sciences, University of Chinese Academy of Sciences},
    addressline={No.19A Yuquan Road},
    city={Beijing},
    postcode={100049},
    country={China}}
\affiliation[4]{organization={School of Fundamental Physics and Mathematical Sciences, Hangzhou Institute for Advanced Study and ICTP-AP, UCAS},
    city={Hangzhou},
    postcode={310024},
    country={China}}
\affiliation[5]{organization={Peng Huanwu Center for Fundamental Theory},
    city={Hefei, Anhui},
    postcode={230026},
    country={China}}

\begin{abstract}
We propose a new splitting behavior of tree-level string/particle amplitudes for massless scalars, gluons and gravitons. We identify certain subspaces in the space of Mandelstam variables, where the universal Koba-Nielsen factor splits into two parts (each with an off-shell leg). Both open- and closed-string amplitudes with Parke-Taylor factors naturally factorize into two stringy {\it currents}, which implies the splitting of bi-adjoint $\phi^3$ amplitudes and via a simple deformation to unified stringy amplitudes, the splitting of amplitudes in the non-linear sigma model and Yang-Mills-scalar theory; the same splitting holds for scalar amplitudes without color such as the special Galileon. Remarkably, if we impose similar constraints on Lorentz products involving polarizations, gluon and graviton amplitudes in bosonic string and superstring theories also split into two (stringy) currents. A special case of the splitting implies soft theorems, and more generally it extends recently proposed smooth splittings and new factorizations near zeros to all these theories. 
\end{abstract}

\maketitle

\section{Introduction}
Perhaps the most familiar property of scattering amplitudes of particles and strings is that on any physical pole, the residue of tree-level amplitudes factorizes into the product of two lower-point ones where an on-shell particle/excitations of strings is exchanged. Recently, two new types of ``factorizing" behavior of scattering amplitudes were observed without going on any physical poles; certain scalar amplitudes simply splits/factorizes into {\it three} parts, when Mandelstam variables are constrained (but no residue is taken). The first is called ``smooth splitting''~\cite{Cachazo:2021wsz} where scalar amplitudes in various theories split into three {\it currents} (each with an off-shell leg), and the second one~\cite{Arkani-Hamed:2023swr} states that color-ordered stringy amplitudes of Tr $\phi^3$, the non-linear sigma model (NLSM) and Yang-Mills-scalar theory (YMS) all factorize into three pieces including a four-point function, which in turn explains their hidden {\it zeros} (also observed for dual resonant amplitudes in the early days of string theory~\cite{DAdda:1971wcy}). As far as we know, the former has been proposed for scalar amplitudes only (but not for particles with spin or strings), and the latter applies to amplitudes with color ordering\footnote{Shortly after the appearance of the first preprint version of this letter, \cite{Bartsch:2024amu} and \cite{Li:2024qfp} appear and discuss zeros and factorization of non-ordered scalar amplitudes. The latter also discuss spinning amplitudes.}.

The fact that these two vastly different approaches demonstrate similar factorization behavior suggests the presence of a profound underlying principle governing this splitting phenomenon. It also raises a question about the generality of such splitting behaviors: do other theories, whether with or without color ordering/spin, manifest analogous splitting behavior after imposing specific constraints on the Mandelstam variables?

In this letter, we propose a new ``splitting'' behavior for tree-level scattering amplitudes of massless scalars, gluons and gravitons (including their string completions, with or without colors), which we call ``2-split'': by restricting Mandelstam variables to a subspace, the amplitude factorizes into the product of two currents. We will see that the key for such a universal behavior lies at the splitting of the Koba-Nielsen factor into two, and provided the splitting of any string correlator under similar conditions on other data such as color or polarizations, the corresponding string amplitude (and particle amplitudes as low-energy limit) must split as expected. 

Interestingly, we will see that our $2$-split provide a common origin for the $3$-split of~\cite{Cachazo:2021wsz} and the factorization near zeros of~\cite{Arkani-Hamed:2023swr}, and also generalize them to a wider context. In particular, the $2$-splitting directly applies to both bosonic and supersymmetric string amplitudes of gluons and gravitons, if we restrict Lorentz products involving their polarizations similar to Mandelstam variables. We will present in~\cite{toappear} that for all these particle amplitudes (and in particular the special Galileon (sGal) for which we do not have a natural stringy completion~\cite{Cachazo:2014xea}), their $2$-split can also be shown directly from scattering equations~\cite{Cachazo:2013gna,Cachazo:2013hca,Cachazo:2013iea} (see~\cite{Naculich:2015zha}), which are saddle-point equations of Koba-Nielsen factor thus inherit the splittings. Last but not least, a special $2$-split implies Weinberg's soft theorems for gluons and gravitons~\cite{Weinberg:1965nx}, just as those ``skinny" zeros of~\cite{Arkani-Hamed:2023swr} implies Adler zeros for Goldstone scalars~\cite{Adler:1964um}. We present examples of the splittings and factorizations near zeros in the appendix. 

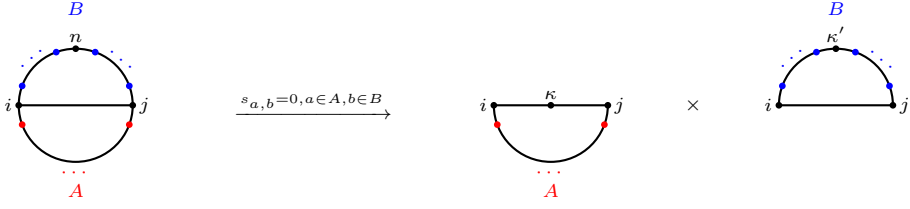
\begin{figure*}[!htbp]
    \centering
\begin{tikzpicture}[scale=1.25]
	\draw[thick] (0,0) circle (0.6);
	\node (p) at (0,0) {};
	\draw (180:0.6) node[left=-1pt]{\scriptsize $i$};
	\draw (0:0.6) node[right=-1pt]{\scriptsize $j$};
	\draw (90:0.6) node[above=-1pt]{\scriptsize $n$};
	\draw[thick] (0:0.6)--(180:0.6);
	\fill (0:0.6) circle (1pt);	\fill (180:0.6) circle (1pt);
	\fill (90:0.6) circle (1pt);

	\fill[blue] (110:0.6) circle (1pt);	
	\fill[blue] (160:0.6) circle (1pt);
	
	\fill[blue] (70:0.6) circle (1pt);	
	\fill[blue] (20:0.6) circle (1pt);
	\draw (45:0.5) node {\scriptsize \begin{rotate}{4}$\color{blue}  \ddots $\end{rotate}};
	\draw (90:0.75) node[above=4pt]{\scriptsize $\color{blue} B$};
	\draw (135:0.5) node {\scriptsize \begin{rotate}{85}$\color{blue}  \ddots $\end{rotate}};
	
	\fill[red] (-20:0.6) circle (1pt);	
	\fill[red] (200:0.6) circle (1pt);	
	\draw (-90:0.6) node[below=5pt]{\scriptsize $\color{red} A$};
	\draw (-90:0.6) node[below=-1pt]{\scriptsize $\color{red} \cdots$};

	\node at (2.5,0) {\scriptsize	$\xrightarrow[]{s_{a,b}=0,a\in A,b\in B}$};

	\node (p1) at (5,0) {};
	
	\draw[thick] ($(p1)+(-0.6,0)$) arc (-180:0:0.6);
	\draw[thick] ($(p1)+(0:0.6)$)--($(p1)+(180:0.6)$);

	\draw ($(p1)$) node[above=-1pt]{\scriptsize $\kappa$};
	\draw ($(p1)+(180:0.6)$) node[left=-1pt]{\scriptsize $i$};
	\draw ($(p1)+(0:0.6)$) node[right=-1pt]{\scriptsize $j$};

	\fill[red] ($(p1)+(-20:0.6)$) circle (1pt);	
	\fill[red] ($(p1)+(200:0.6)$) circle (1pt);	
	\draw ($(p1)+(-90:0.6)$) node[below=5pt]{\scriptsize $\color{red} A$};
	\draw ($(p1)+(-90:0.6)$) node[below=-1pt]{\scriptsize $\color{red} \cdots$};
	
	\fill ($(p1)+(0:0.6)$) circle (1pt);	\fill ($(p1)+(180:0.6)$) circle (1pt);
	\fill ($(p1)$) circle (1pt);

	\node at (6.5,0) {\scriptsize	$\times$};
	
	\node (p2) at (8,0) {};
	
	\draw[thick] ($(p2)+(-0.6,0)$) arc (180:0:0.6);
	\draw[thick] ($(p2)+(0:0.6)$)--($(p2)+(180:0.6)$);
	
	\draw ($(p2)+(90:0.6)$) node[above=-1pt]{\scriptsize $\kappa'$};
	\draw ($(p2)+(180:0.6)$) node[left=-1pt]{\scriptsize $i$};
	\draw ($(p2)+(0:0.6)$) node[right=-1pt]{\scriptsize $j$};

	\fill ($(p2)+(0:0.6)$) circle (1pt);	\fill ($(p2)+(180:0.6)$) circle (1pt);
	\fill ($(p2)+(90:0.6)$) circle (1pt);
	
	\fill[blue] ($(p2)+(110:0.6)$) circle (1pt);	
	\fill[blue] ($(p2)+(160:0.6)$) circle (1pt);
	
	\fill[blue] ($(p2)+(70:0.6)$) circle (1pt);	
	\fill[blue] ($(p2)+(20:0.6)$) circle (1pt);

	\draw ($(p2)+(45:0.5)$) node {\scriptsize \begin{rotate}{4}$\color{blue}  \ddots $\end{rotate}};
	\draw ($(p2)+(90:0.75)$) node[above=4pt]{\scriptsize $\color{blue} B$};
	\draw ($(p2)+(135:0.5)$) node {\scriptsize \begin{rotate}{85}$\color{blue}  \ddots $\end{rotate}};
\end{tikzpicture}
    \caption{$2$-split of a disk: red and blue dots denote particles in $A$ and $B$ respectively (with $A\cup B=[n]/\{i,j,n\}$); by imposing $s_{a\in A, b\in B}=0$ the $n$-particle system ``splits" into $\{i, j\} \cup A$ (and an off-shell $\kappa$), and $\{i,j\}\cup B$ (and an off-shell $\kappa'$). For open string or ordered particle amplitudes, the labels in two subsets preserve the ordering $\kappa,\kappa'$ inserted at the position of $n$; Similar splitting applies to closed string and unordered amplitudes where the ordering on the disk is irrelevant. }
    \label{fig:2fac}
\end{figure*}

\section{Splitting the scattering potential}
Throughout this paper, we work in sufficiently large spacetime dimension ({\it e.g.} with $D>n$), such that the Gram determinant constraints can be ignored. We define {\it 2-split} kinematics as follows: pick $3$ particles, $i,j,k$, and divide the remaining legs into two sets $A, B$, {\it i.e.} $A\cup B=\{1,\cdots, n\}/\{i,j,k\}$, then we demand
\begin{align}
 \label{eq_splitKin}
    s_{a,b} = 0,\qquad \forall a \in A, b \in B\,,
\end{align} 
where $s_{a,b}:=(p_a+p_b)^2=2 p_a \cdot p_b$ (for massless momenta). Without loss of generality, we will choose $k=n$, and for $i <j -1$, $A=(i,j):=\{i{+}1, \cdots, j{-}1\}, B=(j,n)\cup (n,i):=\{j{+}1, \cdots, n{-}1, 1, \cdots, i{-}1\}$; any general 2-split kinematics can be obtained by relabelling.

In order to see how string amplitudes (and their field-theory limits) split under~\eqref{eq_splitKin}, we study the {\it scattering potential}, or logarithm of the Koba-Nielsen factor~\cite{Koba:1969rw}:
\begin{equation}
{\cal S}_n=\sum_{a<b} s_{a,b} \log z_{a,b}=\sum_{a<b\neq k, (a,b) \neq (i,j)} s_{a,b} \log |ab|   
\end{equation}
where $z_{a,b}:=z_b-z_a$; in the second equality we have solved $s_{a,k}$ for $a\neq k$ as well as $s_{i,j}$ in terms of the remaining $n(n{-}3)/2$ independent $s_{a,b}$, and we have defined the SL$(2)$ invariant: $|ab|:=\frac{z_{a, b} z_{i, k} z_{j, k}}{z_{a, k} z_{b, k} z_{i, j}}$. It is convenient to fix the SL$(2)$ redundancy with $z_k\to \infty$ and $z_i=0, z_j=1$, then $|ab|=z_{a,b}$. It is straightforward to see that with~\eqref{eq_splitKin} the potential splits into ``left" and ``right" parts:
\begin{equation}
{\cal S}_n \to \underbrace{({\cal S}_A + {\cal S}_{i, A} + {\cal S}_{j, A})}_{{\cal S}_L(i,A, j; \kappa)}+ \underbrace{({\cal S}_B + {\cal S}_{i, B} +{\cal S}_{j, B})}_{{\cal S}_R(j,B,i; \kappa')}\,,    
\end{equation}
where ${\cal S}_A=\sum_{a<b, a,b \in A} s_{a,b} \log |ab|$, ${\cal S}_{i, A}=\sum_{a\in A} s_{i,a} \log |i a|$, ${\cal S}_{j, A}=\sum_{a\in A} s_{a, j} \log |a j|$ (in the gauge fixing above, $|ia|=z_a-z_i=z_a$, $|aj|=z_j-z_a=1-z_a$) and similarly for the right part; in the second equality we have interpreted them as the scattering potential for two currents: the first one with on-shell legs $a\in A$, $i, j$, and the second one with $b\in B, i, j$, and each of them contains an off-shell leg with momentum
$p_\kappa=-\sum_{a\in A} p_a -p_i-p_j$, and $p_{\kappa'}=-\sum_{b\in B} p_b-p_i-p_j$, respectively~\footnote{The left/right currents contain $|A|+3$ and $|B|+3$ external legs respectively, and in total we have $n+3$ legs; the dimensions of these moduli spaces add up: $n{-}3=|A|+|B|=n_L{-}3+n_R{-}3$.}~(see Figure~\ref{fig:2fac}). Note that for the left/right currents, we have fixed $z_i=0, z_j=1$ and both $z_\kappa, z_\kappa' \to \infty$, which breaks the symmetry between $i,j$ and $k$; in other words, we have chosen $i,j$ to be on-shell in both currents, which means the remaining $\kappa/\kappa'$ (replacing leg $k$) must be off-shell, and we could equally make other choices.

For $n$-point open-string amplitude, we define the measure including Koba-Nielsen factor as $d\mu_n^{\mathbb{R}}:= (\alpha')^{n{-}3} \prod_{a \neq i,j,k} d z_a \exp(\alpha' {\cal S}_n)$, where the SL(2) redundancy is fixed {\it e.g.} $z_i=0, z_j=1$ and $z_k\to \infty$. 
We conclude that the measure factorizes:
\begin{equation}
d\mu^{\mathbb{R}
}_n \to d \mu^{\mathbb{R}
}_L (i, A, j; \kappa) d\mu^{\mathbb{R}
}_R (j, B, i; \kappa')
\,.  
\end{equation}
and similarly the closed-string measure $d\mu_n^{\mathbb{C}}=d\mu_n^{\mathbb{R}}(z) d\mu_n^{\mathbb{R}}(\bar{z})$ also factorizes into $L$ and $R$ parts. 

Before proceeding, we show that for the scattering potential, both the $3$-split of~\cite{Cachazo:2021wsz} and factorizations near zeros of~\cite{Arkani-Hamed:2023swr} follow from this basic $2$-split. For the former, let us call $B$ as $B'$ instead (assume $|B|>1$); we further split it as $B'=B\cup C$ and demand $s_{b,c}=0$ for $b\in B, c\in C$, then the scattering potential splits into three:
\begin{equation}
{\cal S}_n \to {\cal S} (i, A, j; \kappa_A) + {\cal S}(j, B, k; \kappa_B)+{\cal S}(k, C, i; \kappa_C)    
\end{equation}
with off-shell momenta of $\kappa_{A}, \kappa_B, \kappa_C$ given by momentum conservation. 
We remark that it is such $3$-split that deserves the name ``smooth splitting" since all $n$ on-shell legs appear in the currents (with $i,j,k$ each shared by two of them and the symmetry between $i,j,k$ restored); note that our special (``skinny'')  case where we have {\it e.g.} $|A|=1$ corresponds to the special case of~\cite{Cachazo:2021wsz} where one of the three currents becomes the trivial $3$-point one.

For the latter, note that in the special case when $A$ has only one particle, {\it e.g.} $m$, \eqref{eq_splitKin} corresponds to the factorization near ``skinny" zero of~\cite{Arkani-Hamed:2023swr}: the amplitude factorizes into the $(n{-}1)$-point current (with on-shell legs $[n]/{k,m}$ and the off-shell leg $\kappa$), times a $4$-point function (with on-shell legs $i,j$ and two more off-shell legs), as well as a trivial $3$-point current. In general to see zeros and factorizations we further set $s_{a,k}=0$ for all $a\in A$ {\it except for $a=m$}, and the left-potential further splits
\begin{equation}
{\cal S}_L(i, A, j; \kappa)\to {\cal S}_L(i, A/\{m\}, j; \rho) + {\cal S}(i, \rho', j, \kappa),
\end{equation}
into a current without leg $m$ and a {\it four-point} potential with two off-shell legs (alternatively we have a further splitting of ${\cal S}_R(j, B, i; \kappa')$ if we choose $s_{b,k}=0$ for all $b\in B$ except for $b=m$ for some $m\in B$).
In deriving this we have used $s_{a,\kappa}=0$ for $a \in A$ and $a\neq m$, and note $p_\rho=
-\sum_{a\neq m} p_a-p_i-p_j,p_{\rho'}=
\sum_{a\in A} p_{a}$.  

\begin{figure*}[!htbp]
    \centering
    \includegraphics[scale=0.45]{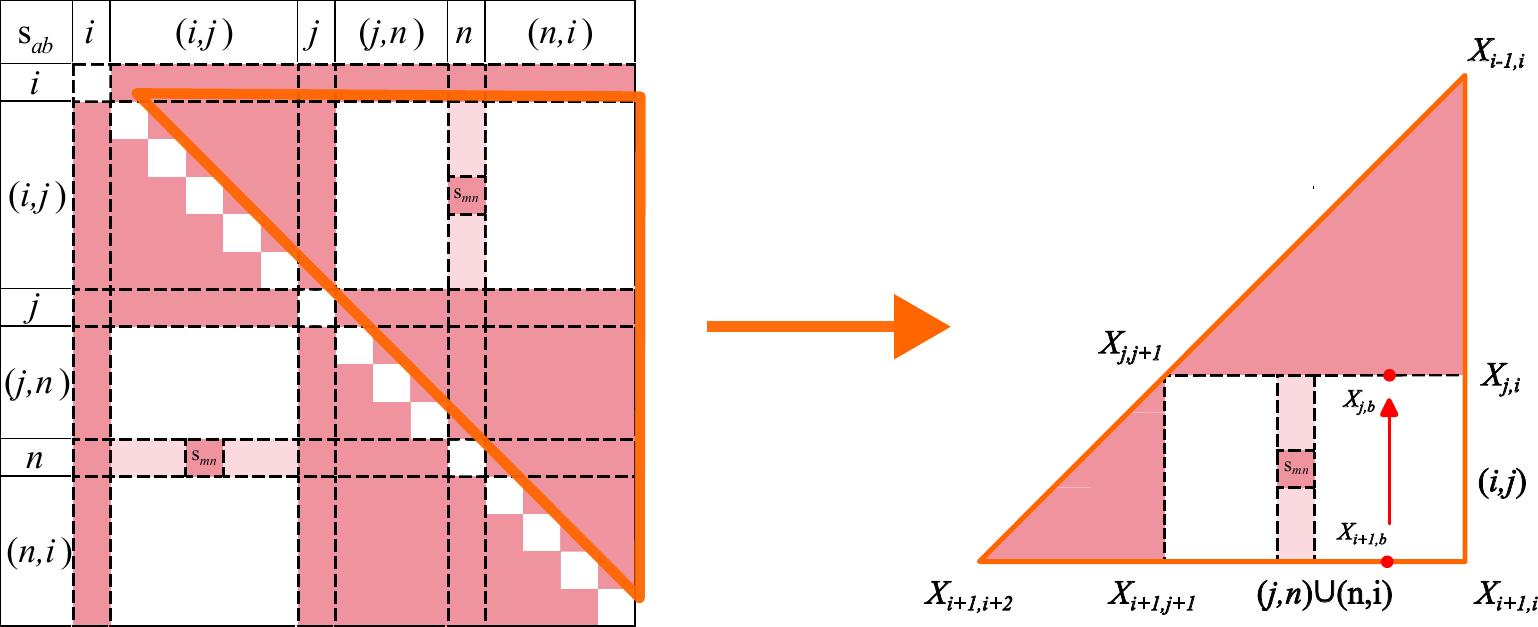}
    \caption{The Mandelstam matrix (left) and the kinematic mesh (right) for $2$-split kinematics: $s_{a,b}=0$ are denoted by blank entries, and the additional $s_{a,n}=0$ (except for $a=m$) are denoted by light pink entries, with $a \in A=(i,j):=\{i{+}1, \cdots, j{-}1\}, b\in B = (j,n)\bigcup(n,i) = \{j{+}1, \cdots, n{-}1, 1, \cdots, i{-}1\} $. We define $X_{i,j} = (p_i + \dots +p_{j{-}1})^2$. Under the factorization,  $X_{i,j}$ variables will be shifted as indicated by the arrow in the mesh (right), \textit{c.f.} \cite{Arkani-Hamed:2023swr}. }
    \label{fig_ManMatrix2mesh}
\end{figure*}

Very nicely, this is nothing but the factorizations near zeros of~\cite{Arkani-Hamed:2023swr}: at least for scalars the four-point ``amplitude" vanishes when we finally set $s_{m,k}=0$, and the relationship between the Mandelstam matrix and the kinematic mesh is illustrated in Figure \ref{fig_ManMatrix2mesh}: the zeros of color-ordered amplitudes of Tr$\phi^3$, NLSM or YMS theory correspond to $s_{a,b}=0$ for $i<a<j$ and $j<b\leq n$ (including $b=n$) and $1\leq b<i$, which is precisely given by a ``rectangle" in the mesh picture of associahedron~\cite{Arkani-Hamed:2017mur, Arkani-Hamed:2019vag}, and by excluding $s_{m, n}=0$ for some $i<m<j$ we recover the factorization near zero in the mesh. The currents can be written as the same functions as amplitudes, where the planar variables are exactly shifted as in~\cite{Arkani-Hamed:2023swr} due to the off-shell leg such as $p_{\rho}^2 \neq 0$. See the appendix for a brief review of factorizations near zeros.

\section{Splitting scalar amplitudes}
\paragraph{Stringy $\phi^3$, NLSM and YMS} Let us begin with the splitting of scalar amplitudes, and first show how open- and closed-string integrals that reduce to bi-adjoint $\phi^3$ amplitudes in $\alpha'\to 0$ limit (known in the literature~\cite{Mafra:2011nw, Carrasco:2016ldy,Schlotterer:2018zce,Stieberger:2014hba,Stieberger:2013wea,Broedel:2013tta} as $Z$ and $J$ integrals respectively) split under \eqref{eq_splitKin}. Both integrals depend on two orderings $\alpha, \beta$:
\begin{equation}
Z_{\alpha|\beta}:=\int_{D(\alpha)} d\mu_n^{\mathbb{R}}~{\rm PT}(\beta)\,,\quad J_{\alpha, \beta}:=\int d\mu_n^{\mathbb{C}}~{\rm PT}(\alpha){\rm PT}(\beta)\,,    
\end{equation}
where for canonical ordering $I=(1,2,\cdots, n)$, the integration domain for $Z$ integral reads $D(I):=\{z_1<z_2<\cdots <z_{n{-}1}\}$ (with $z_n\to \infty$ and {\it e.g.} $z_i=0, z_j=1$ fixed), and the gauge-fixed Parke-Taylor factor is defined as 
\begin{equation}
{\rm PT}(I):=\frac{\Delta_{i,j,n}}{z_{1,2} z_{2,3} \cdots z_{n,1}}\,,
\end{equation}
with the Jacobian from gauge fixing $\Delta_{i,j,k}:=z_{i,j} z_{j,k} z_{k,i}$ (note that all dependence on $z_n\to \infty$ cancel), and similarly for any orderings $\alpha, \beta$. The splitting only happens for orderings that are {\it compatible} with the $2$-split kinematics; without loss of generality we fix $\alpha=(12 \cdots n)$, which is compatible with our chosen kinematics above and in general compatible orderings correspond to permutations of that act on sets $A=(i,j), B\cup \{n\}=(j,i)$, {\it i.e.} $\beta=(i, \beta_{(i,j)}, j, \beta_{(j, i)})$. After gauge fixing, it is easy to see that any such Parke-Taylor factor splits nicely by including Jacobian factor $\Delta_{i,j,\kappa}$ and $\Delta_{i,j, \kappa'}$ on the RHS:
\begin{equation}
{\rm PT}(I) = {\rm PT}(i,(i,j),j,\kappa)\times  {\rm PT}(j, (j,n), \kappa', (n,i), i)\,,
\end{equation} 
which, together with the splitting of $D(\alpha)$, imply that the $Z$ integral splits into two (stringy) currents:
\begin{equation}
Z_{\alpha|\beta} \to Z_{\alpha|\beta}(i,A,j; \kappa) \times Z_{\alpha|\beta} (j,B,i; \kappa')   
\end{equation}
where $\alpha, \beta$ acts on corresponding legs in $A, B$ and we emphasize that $\kappa, \kappa'$ are off-shell legs. Exactly the same holds for closed-string $J$ integrals. 

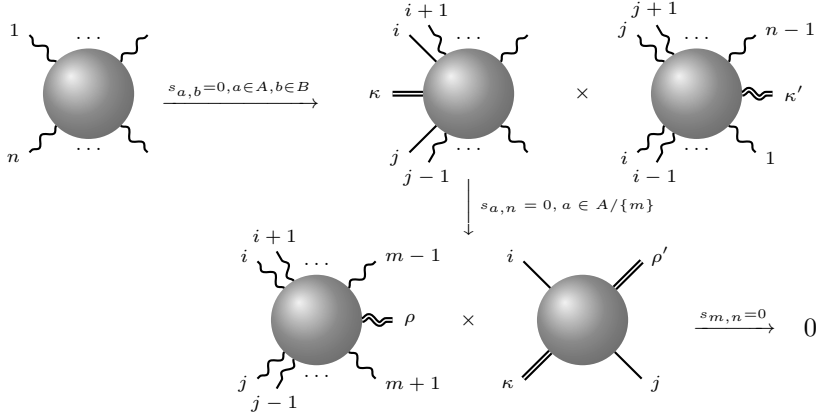
\begin{figure*}[!htbp]
    \centering
\begin{tikzpicture}
	\shade[ball color = gray, opacity=0.7] (0,0) circle (0.6);
	\node (p) at (0,0) {};
	\draw[thick,decorate,decoration=snake, segment length=8pt, segment amplitude=1.5pt] (45:0.6)--++(45:0.5) ;
	\draw[thick,decorate,decoration=snake, segment length=8pt, segment amplitude=1.5pt] (-45:0.6)--++(-45:0.5) ;
	\draw[thick,decorate,decoration=snake, segment length=8pt, segment amplitude=1.5pt] (135:0.6)--++(135:0.5) ;
	\draw[thick,decorate,decoration=snake, segment length=8pt, segment amplitude=1.5pt] (-135:0.6)--++(-135:0.5) ;
	\draw (135:1.2) node[left=-2pt]{\scriptsize $1$};
	\draw (-135:1.2) node[left=-2pt]{\scriptsize $n$};
	\draw (90:0.6) node[above=0pt]{\scriptsize $\ldots$};
	\draw (270:0.6) node[below=0pt]{\scriptsize $\ldots$};

	\node at (2,0) {\scriptsize	$\xrightarrow[]{s_{a,b}=0,a\in A,b\in B}$};

	\shade[ball color = gray, opacity=0.7] (5,0) circle (0.6);
	\node (p1) at (5,0) {};
	\draw[thick,decorate,decoration=snake, segment length=8pt, segment amplitude=1.5pt] ($(p1)+(45:0.6)$)--++(45:0.5) ;
	\draw[thick,decorate,decoration=snake, segment length=8pt, segment amplitude=1.5pt] ($(p1)+(-45:0.6)$)--++(-45:0.5) ;
	\draw[thick] ($(p1)+(135:0.6)$)--++(135:0.5) ;
	\draw[thick] ($(p1)+(-135:0.6)$)--++(-135:0.5) ;
	\draw[thick,double] ($(p1)+(180:0.6)$)--++(180:0.4) ;
	\draw ($(p1)+(180:1.1)$) node[left=-2pt]{\scriptsize $\kappa$};
	\draw ($(p1)+(135:1.2)$) node[left=-2pt]{\scriptsize $i$};
	\draw ($(p1)+(-135:1.2)$) node[left=-2pt]{\scriptsize $j$};
	\draw ($(p1)+(90:0.6)$) node[above=0pt]{\scriptsize $\ldots$};
	\draw ($(p1)+(270:0.6)$) node[below=0pt]{\scriptsize $\ldots$};

    \draw[thick,decorate,decoration=snake, segment length=8pt, segment amplitude=1.5pt] ($(p1)+(120:0.6)$)--++(120:0.45) ;
    \draw[thick,decorate,decoration=snake, segment length=8pt, segment amplitude=1.5pt] ($(p1)+(-120:0.6)$)--++(-120:0.45) ;
    \draw ($(p1)+(120:1.1)$) node[above=-2pt]{\scriptsize $i+1$};
    \draw ($(p1)+(-120:1.1)$) node[below=-2pt]{\scriptsize $j-1$};

	\node at (6.5,0) {\scriptsize	$\times$};

	\shade[ball color = gray, opacity=0.7] (8,0) circle (0.6);
	\node (p2) at (8,0) {};
	\draw[thick,decorate,decoration=snake, segment length=8pt, segment amplitude=1.5pt] ($(p2)+(45:0.6)$)--++(45:0.5) ;
	\draw[thick,decorate,decoration=snake, segment length=8pt, segment amplitude=1.5pt] ($(p2)+(-45:0.6)$)--++(-45:0.5) ;
	\draw[thick,decorate,decoration=snake, segment length=8pt, segment amplitude=1.5pt] ($(p2)+(135:0.6)$)--++(135:0.5) ;
	\draw[thick,decorate,decoration=snake, segment length=8pt, segment amplitude=1.5pt] ($(p2)+(-135:0.6)$)--++(-135:0.5) ;
	\draw[thick,double,decorate,decoration=snake, segment length=8pt, segment amplitude=1.5pt] ($(p2)+(0:0.6)$)--++(0:0.4) ;
	\draw ($(p2)+(0:1.1)$) node[right=-2pt]{\scriptsize $\kappa^\prime$};
	\draw ($(p2)+(135:1.2)$) node[left=-2pt]{\scriptsize $j$};
	\draw ($(p2)+(-135:1.2)$) node[left=-2pt]{\scriptsize $i$};
	\draw ($(p2)+(45:1.2)$) node[right=-2pt]{\scriptsize $n-1$};
	\draw ($(p2)+(-45:1.2)$) node[right=-2pt]{\scriptsize $1$};
	\draw ($(p2)+(90:0.6)$) node[above=0pt]{\scriptsize $\ldots$};
	\draw ($(p2)+(270:0.6)$) node[below=0pt]{\scriptsize $\ldots$};

    \draw[thick,decorate,decoration=snake, segment length=8pt, segment amplitude=1.5pt] ($(p2)+(120:0.6)$)--++(120:0.45) ;
    \draw[thick,decorate,decoration=snake, segment length=8pt, segment amplitude=1.5pt] ($(p2)+(-120:0.6)$)--++(-120:0.45) ;
    \draw ($(p2)+(120:1.1)$) node[above=-2pt]{\scriptsize $j+1$};
    \draw ($(p2)+(-120:1.1)$) node[below=-2pt]{\scriptsize $i-1$};

	\node at (6.3,-1.5) {\tiny $ s_{a,n}=0,a\in A/\{m\}$};
	\node at (5,-1.5) {\scriptsize $\Bigg\downarrow$};

	\node at (5,-3) {\scriptsize $ \times$};
	\shade[ball color = gray, opacity=0.7] (6.5,-3) circle (0.6);
	\node (q1) at (6.5,-3) {};
	\draw[thick,double] ($(q1)+(45:0.6)$)--++(45:0.5) ;
	\draw[thick] ($(q1)+(-45:0.6)$)--++(-45:0.5) ;
	\draw[thick] ($(q1)+(135:0.6)$)--++(135:0.5) ;
	\draw[thick,double] ($(q1)+(-135:0.6)$)--++(-135:0.5) ;
	\draw ($(q1)+(135:1.2)$) node[left=-2pt]{\scriptsize $i$};
	\draw ($(q1)+(45:1.2)$) node[right=-2pt]{\scriptsize $\rho^\prime$};
	\draw ($(q1)+(-45:1.2)$) node[right=-2pt]{\scriptsize $j$};
	\draw ($(q1)+(-135:1.2)$) node[left=-2pt]{\scriptsize $\kappa$};
	
	\shade[ball color = gray, opacity=0.7] (3,-3) circle (0.6);
	\node (q2) at (3,-3) {};
	\draw[thick,decorate,decoration=snake, segment length=8pt, segment amplitude=1.5pt] ($(q2)+(45:0.6)$)--++(45:0.5) ;
	\draw[thick,decorate,decoration=snake, segment length=8pt, segment amplitude=1.5pt] ($(q2)+(-45:0.6)$)--++(-45:0.5) ;
	\draw[thick,decorate,decoration=snake, segment length=8pt, segment amplitude=1.5pt] ($(q2)+(135:0.6)$)--++(135:0.5) ;
	\draw[thick,decorate,decoration=snake, segment length=8pt, segment amplitude=1.5pt] ($(q2)+(-135:0.6)$)--++(-135:0.5) ;
	\draw[thick,double,decorate,decoration=snake, segment length=8pt, segment amplitude=1.5pt] ($(q2)+(0:0.6)$)--++(0:0.4) ;
	\draw ($(q2)+(0:1.1)$) node[right=-2pt]{\scriptsize $\rho$};
	\draw ($(q2)+(135:1.2)$) node[left=-2pt]{\scriptsize $i$};
	\draw ($(q2)+(-135:1.2)$) node[left=-2pt]{\scriptsize $j$};
	\draw ($(q2)+(45:1.2)$) node[right=-2pt]{\scriptsize $m-1$};
	\draw ($(q2)+(-45:1.2)$) node[right=-2pt]{\scriptsize $m+1$};
	\draw ($(q2)+(90:0.6)$) node[above=0pt]{\scriptsize $\ldots$};
	\draw ($(q2)+(270:0.6)$) node[below=0pt]{\scriptsize $\ldots$};

    \draw[thick,decorate,decoration=snake, segment length=8pt, segment amplitude=1.5pt] ($(q2)+(120:0.6)$)--++(120:0.45) ;
    \draw[thick,decorate,decoration=snake, segment length=8pt, segment amplitude=1.5pt] ($(q2)+(-120:0.6)$)--++(-120:0.45) ;
    \draw ($(q2)+(120:1.1)$) node[above=-2pt]{\scriptsize $i+1$};
    \draw ($(q2)+(-120:1.1)$) node[below=-2pt]{\scriptsize $j-1$};

    \node at (8.5,-3) {\scriptsize	$\xrightarrow[]{s_{m,n}=0}$ } ;
    \node at (9.5,-3.1) {	$0$ } ;
 
\end{tikzpicture}
	
    \caption{The splitting of an amplitude with $n$ legs (denoted by wavy lines) into a ``mixed" currents with $3$ $\phi$'s (denoted by straight lines) and a ``pure" current off-shell legs denoted by double lines); on the second line we show a further ``splitting" of a scalar current (such as in NLSM/sGal), which leads to a factorization near the zero contained in the four-point function. 
    }
    \label{fig:2facAnd3fac}
\end{figure*}

In~\cite{Arkani-Hamed:2023swr}, a simple deformation of the diagonal $Z$-integral ($\beta=\alpha = I$) was introduced to give a unified tree-level stringy amplitude in Tr$\phi^3$, NLSM and YMS with even $n$ (see~\cite{Arkani-Hamed:2023jry, Arkani-Hamed:2024nhp} for loops). The deformed stringy Tr$\phi^3$ amplitude is defined as
\begin{align}
&Z_n^\delta:=\int_{D(I)}d\mu_n^{\mathbb{R}}~{\rm PT}^{\delta}(I)\,,\\\nonumber
&{\rm PT}^\delta (I)
:=\Delta_{i,j,n} z_{1,2}^{-1-\delta} z_{2,3}^{-1+\delta} \cdots z_{n,1}^{-1+\delta}\,, 
\end{align}
where $\delta \in \mathbb{R}$ is the deform parameter.
Very nicely, we find that such deformed stringy amplitude splits almost exactly as before:
\begin{equation}
Z_n^\delta\to Z_L^{(-)^i \delta} (i, A, j; \kappa)  Z_R^{(-)^j \delta}(j,B,i; \kappa')\,,   
\end{equation}
where on the left/right current the deformation parameter is $\pm \delta$ depending on $i/j$ being even or odd. Note that $|A|+|B|=n{-}3$ is odd, thus one current has even multiplicity and the other has odd. In the $\alpha'\to 0$ limit, $Z_n^\delta$ with any generic (non-integer) $\delta$ gives NLSM amplitude, while for $\delta=1$ it gives YMS amplitude with pairs of scalars $(12)(34) \cdots (n{-}1 n)$ ($\delta=-1$ gives the amplitude with pairs $(23)(45)\cdots (n1)$). Therefore, the NLSM/YMS amplitude splits into an NLSM/YMS current (with even multiplicity) and a ``mixed" current with $3$ $\phi$'s~\cite{Cachazo:2016njl}. For example, for $|A|$ being odd, we have (see Figure~\ref{fig:2facAnd3fac})
\begin{equation}
\label{eq_NLSM2split}
{\cal M}^{\rm NLSM}_n\to {\cal J}^{\rm NLSM} (i, A, j; \kappa) {\cal J}^{\rm mixed} (j^\phi, B, i^\phi; \kappa'^{\phi})\,,
\end{equation}
and for YMS one needs to be careful about where $i,j$ {\it etc.} appear in the scalar pairs~\cite{Cao:2024qpp}.

Moreover, we recover the factorizations near zeros for the unified stringy amplitude of~\cite{Arkani-Hamed:2023swr} if we further pick an arbitrary particle $m \in A$, and set {\it e.g.} $s_{a,n}=0$ for $a\in A/\{m\}$. The amplitude factorizes into two currents with $|A|{+}2$ and $|B|{+}3$ legs ($|B|=n{-}3{-}|A|$), times a four-point function; crucially there are two possibilities with $|A|$ either even or odd. For the former ({\it e.g.} with $i$ even and $j$ odd), we have two stringy NLSM/YMS currents times a Beta function $B(s,t):=\Gamma(s)\Gamma(t)/\Gamma(s+t)$ (with $\alpha'$ suppressed):
\begin{equation}
Z_n^\delta \to Z^{\delta}(i, A/\{m\}, j; \rho) B(s_{i, \kappa}, s_{j, \kappa}) Z^{-\delta}(j, B, i; \kappa')\,,
\end{equation}
where $s_{i,\kappa}:=(p_i+p_\kappa)^2=(\sum_{a\in A}p_a+p_j)^2$. For the latter ({\it e.g.} with $i,j$ even), we have two stringy mixed currents with $3$ $\phi$'s, times a shifted Beta function:
\begin{equation}
Z_n^\delta \to
Z^{\delta}(i, A/\{m\}, j; \rho) B(s_{i, \kappa} - \delta, s_{j, \kappa} +\delta)
Z^{\delta}(j, B, i; \kappa')\,.
\end{equation}
By \eqref{eq_splitKin} we have $s_{i,\kappa}+s_{j,\kappa}= -s_{m,n}$, thus by setting $s_{m,n}=0$ (or positive integer), $1/\Gamma(s_{i,\kappa}+s_{j,\kappa})$ vanishes which reproduces zero of the amplitude~\cite{Arkani-Hamed:2023swr}. Furthermore, the $\alpha'\to 0$ limit gives ``pure $\times$ pure" $\times$ ${\cal M}^{\phi^3}_4$, and ``mixed $\times$ mixed" $\times$ ${\cal M}^{\rm NLSM/YMS}_4$, respectively~\cite{Arkani-Hamed:2023swr}; recall that ${\cal M}_4^{\phi^3}=1/s+1/t$, ${\cal M}^{\rm NLSM}_4=s+t$ and ${\cal M}^{\rm YMS}_4=(s+t)/s$.

\paragraph{Scalars without color} Remarkably, we obtain the same splitting for scalars without color, such as those in sGal, Dirac-Born-Infeld (DBI) and Einstein-Maxwell-scalar (EMS) with even $n$. We do not know any simple stringy model similar to $Z^\delta$ for these amplitudes, but the splitting of all field-theory amplitudes (including $\phi^3$, NLSM and YMS, as well as Yang-Mills and gravity amplitudes to be discussed below) can be derived using formulas based on scattering equations~\cite{Cachazo:2013gna, Cachazo:2013iea, Cachazo:2014xea}. As we will discuss in detail in~\cite{toappear}, the splitting of the universal measure including scattering equations follow from that of the scattering potential, and very nicely ``integrands" for these amplitudes (such as ${\rm PT}$ or ${\rm det}' A_n$) split as well! For example, the splitting of ${\rm det}' A_n$ guarantees that not only NLSM but also sGal amplitudes split:   
{\it e.g.} with $|A|$ odd, we obtain for field-theory amplitude(see Figure~\ref{fig:2facAnd3fac}):
\begin{equation}
\label{eq_sGal2split}
{\cal M}^{\rm sGal}_n \to {\cal J}^{\rm sGal} (i, A, j; \kappa) \times {\cal J}^{\rm mixed} (j^\phi, B, i^\phi; \kappa'^\phi)\,.  
\end{equation}
Similarly, the amplitudes in DBI and EMS with appropriate scalar pairs split just as those in YMS. These results in turn imply new zeros and factorizations for these amplitudes without color. The upshot is:
\begin{itemize}
    \item The amplitude vanishes for $s_{a,b}=0$ with $a\in A$ and $b\in B':=B \cup \{k\}$. 
    \item The amplitude factorizes when we turn on $s_{k,m}\neq 0$, for any $m \in A$:
\begin{equation}
{\cal M}_n \to {\cal M}_4 \times {\cal J}(i, A/\{m\}, j; \rho) {\cal J}(j, B, i; \rho')\,.
\end{equation}
\end{itemize}
Let us specify to sGal to be concrete: if both $|A|, |B|$ are even, these are sGal currents times ${\cal M}_4^{\phi^3}$; if both $|A|, |B|$ are odd, these are mixed currents with $3$ $\phi$'s, times ${\cal M}^{\rm sGal}_4=-s t (s+t)$ ($s:=s_{i,\kappa}$, $t:=s_{j,\kappa}$). Similar results hold for DBI and EMS amplitudes.

\section{Splitting string amplitudes of gluons and gravitons}
In this section, we show the splitting of gluon and graviton amplitudes in bosonic string and superstring theories. Factorizations near zeros for Yang-Mills amplitudes have been considered from ``scaffolding" YMS amplitudes in~\cite{Arkani-Hamed:2023swr, Arkani-Hamed:2023jry}, but here we adopt a different approach and find that open- and closed-string amplitudes for gluons and gravitons split if we impose conditions similar to \eqref{eq_splitKin} for polarizations. More details about the derivation for the splitting of string correlators as well as Cachazo-He-Yuan (CHY) formulas will be given in~\cite{toappear}.

Similar to scalar cases, we expect a current with gluons/gravitons only (albeit with one off-shell leg $\kappa'$, which carries the polarization of leg $n$), and a mixed current with $i,j,\kappa$ being scalars. This can be achieved if we impose the following conditions on top of \eqref{eq_splitKin}
\begin{equation}\label{eq_pol}
\epsilon_a \cdot \epsilon_{b'}=0\,, \quad p_a \cdot \epsilon_{b'}=0\,,\quad \epsilon_a \cdot p_b=0\,,
\end{equation}
for $a\in A, b \in B$ and $b' \in B \cup \{i,j,n\}$. Under these conditions, we claim that gluon amplitudes in bosonic string and superstring theory~\cite{Green:1987sp} split as:
\begin{equation}\label{bos_string}
{\cal M}^{\rm open}_n \to {\cal J}^{\rm mixed} (i^\phi, A, j^\phi; \kappa^\phi) \times {\cal J} (j, B, i; \kappa')_\mu \epsilon_n^\mu\,,  
\end{equation}
where the ``pure" gluon current has an off-shell leg $\kappa'$, and it is contracted with polarization $\epsilon_n$(see the first line of Figure~\ref{fig:2facAnd3fac} with wavy lines denote gluons). Note that there is special case with $B=\emptyset$ where \eqref{eq_splitKin} imposes no conditions, but \eqref{eq_pol} turns off Lorentz product between $\epsilon_{i,j,k}$ and $\epsilon_a/p_a$ for the remaining $n{-}3$ legs; in this case ${\cal J}^{\rm mixed}$ is a current with $n{-}3$ gluons (in $A$) and $3$ $\phi's$, while ${\cal J}(i,j;\kappa')$ is given by the familiar $3$-gluon current. 

The proof for \eqref{bos_string} is essentially the same as before: with \eqref{eq_pol} the (gauge-fixed) string correlator ${\cal C}_n(\{\epsilon, p, z\})$ factorizes into a correlator with gluons in $A$ (times a ``Parke-Taylor" factor for $i,j,\kappa$), and one with gluons in $B\cup \{i,j, \kappa'\}$ (with polarization of $\kappa'$ replaced by $\epsilon_n$). We give a derivation of this for bosonic string correlator in the appendix; more details and a similar derivation for superstring correlator will be given in~\cite{toappear}. 

For gravity amplitudes in closed-string theory, the correlator is given by ${\cal C}_n(\epsilon, p, z) {\cal C}_n (\tilde{\epsilon}, p, \bar{z})$ (with polarization tensor $\varepsilon^{\mu \nu}=\epsilon^\mu \tilde{\epsilon}^\nu$), and one can impose conditions \eqref{eq_pol} separately on $\epsilon$ and $\tilde{\epsilon}$ which lead to different splittings; one choice leads to the same splitting as in \eqref{bos_string} and another one yields two mixed currents with $3$ gluons in Einstein-Yang-Mills theory (legs in $A,B$ are gravitons):
\begin{equation}
{\cal M}^{\rm closed}_n \to {\cal J}(i^g, A, j^g; \kappa^g)_\mu \epsilon_n^\mu \times {\cal J} (j^g, B, i^g; {\kappa'}^g)_\nu \tilde{\epsilon}_{n
}^\nu\,.
\end{equation}
One can derive factorization near zeros for gluon amplitudes by further imposing $s_{a,n}=0$ for $a\in A, a\neq m$ (or $s_{b,n}=0$ for $b\in A, b\neq m$) and similarly for Lorentz products with polarizations. There are two possibilities: either we have a pure current, a mixed current with $3$ $\phi$'s times a $4$-point function with $1$ gluon, or two mixed currents times a $4$-gluon function. We obtain zeros of the amplitude from those of the four-point function, which differ from zeros obtained by ``scaffolding" YMS amplitudes~\cite{Arkani-Hamed:2023swr, Arkani-Hamed:2023jry}; the latter correspond to setting $p_a\cdot p_b=p_a \cdot \epsilon_b =\epsilon_a \cdot p_b=\epsilon_a \cdot \epsilon_b=0$ for $a\in A, b \in B \cup \{k\}$. 

\section{Soft theorems} 
We comment on the relation of splitting with the soft theorems for gluons/gravitons~\cite{Weinberg:1965nx} and (enhanced) Adler zeros for scalars~\cite{Adler:1964um,Cheung:2014dqa}. We are interested in the special ``skinny" case with $|A|=1$. The gluon amplitude splits into $(n{-}1)$-point current and a four-point mixed one, which can be computed exactly ${\cal J}^{\rm mixed} (i^\phi, a, j^\phi; \kappa^\phi)=\epsilon_a \cdot p_i B(s_{i,a}, s_{j,a}+1) -\epsilon_a \cdot p_j B(s_{i,a}+1, s_{j,a})$. 

Note that we have only imposed $s_{a,b\in B}=0$ which does not imply the softness of $p_a$; now the soft limit is reached by further imposing $s_{a,i}, s_{a,j}\to 0$ (thus $s_{a,n}=0$), in which case the current becomes an amplitude ($\kappa'$ becomes on-shell). In other words, instead of sending all $s_{a, b'}$ for $b' \in [n]/\{a\}$ to zero simultaneously, we are taking a two-step procedure, and we need to sum over all possible assignments of $i,j,k \neq a$; 
since $i,j$ are fixed to be adjacent to $a$ in the color ordering which are the only contributions at leading order, we only need to sum over $k$ where each term gives identical result, thus up to possible overall constants we obtain
\begin{equation}
 \sum_{k\neq i,j, a} {\cal J}^{\rm mixed}  \times {\cal J}_{n{-}1}
 \to  \left(\frac{\epsilon_a \cdot p_i}{p_a\cdot p_i}{-} \frac{\epsilon_a \cdot p_j}{p_a\cdot p_j}\right) \times {\cal M}_{n{-}1}^{\rm YM},
\end{equation}
where the mixed current simplifies to nothing but the soft gluon factor! Although we have imposed restrictions on the polarizations~\eqref{eq_pol} not needed for soft limit, they do not appear at leading order. A similar argument applies to the soft graviton, where we need to sum over triplets $i,j,k \neq a$ since any mixed current with one graviton and three $\phi$'s contributes to the leading soft factor: 
\begin{equation}
\sum_{k,i,j\neq a}{\cal J}^{\rm mixed} \times {\cal J}_{n{-}1}
\to \left(\sum_{b\neq a} \frac{\epsilon_a \cdot p_b \tilde\epsilon_a \cdot p_b}{p_a \cdot p_b}\right) {\cal M}_{n{-}1}^{\rm GR} 
\end{equation}

As already pointed out in~\cite{Arkani-Hamed:2023swr}, the special ``skinny" zeros in NLSM implies the Adler zero, which also generalizes to enhanced Adler zeros of DBI and sGal since in the soft limit $p_a=\tau \hat{p}_a$ with $\tau \to 0$, their four-point functions behave like ${\cal M}_4 \sim \tau^s$ for $s=1,2,3$, respectively. What multiplies ${\cal M}_4$ is a $n{-}1$-point mixed current with $3$ $\phi$'s, thus we expect that one can derive from our splitting ``the coefficient of Adler zero"~\cite{Cachazo:2016njl}, which involves sum of such mixed currents at least for the NLSM case. 

\section{Outlook}
In this letter we have demonstrated that, in certain kinematic subspaces string amplitudes and particle amplitudes that admit a CHY representation split into a product of two lower-point currents. This explains and extends the recently proposed smooth splitting and factorizations near zeros, and in a sense also generalizes soft theorems. We would like to understand better the relation between ``skinny" splitting and soft theorems, and to study multi-soft limits using more general splittings, similar to what have been considered very recently in~\cite{Arkani-Hamed:2024yvu} for NLSM. 

Again inspired by~\cite{Arkani-Hamed:2024yvu}, it would be highly desirable to generalize such splitting to loop integrands at least in some theories, perhaps via their fascinating connections with ``surfaceology"~\cite{Arkani-Hamed:2023lbd,Arkani-Hamed:2023mvg, Arkani-Hamed:2024vna}. Clearly the splittings ``commute" with double copy~\cite{Kawai:1985xq, Bern:2008qj} in the sense that they apply to both open and closed-string amplitudes, and it would be nice to understand how to see this explicitly from either KLT or BCJ construction. It would be interesting to extend the splittings to (supersymmetric) amplitudes with fermions in specific dimensions, and {\it e.g.} to explore them using spinor-helicity variables. Finally, we would like to understand the nature of all these splitting behavior of scattering amplitudes~\cite{Cachazo:2021wsz, Arkani-Hamed:2023swr, toappear}, especially to see if there is any physical principle behind all of them.

\paragraph{Acknowledgments}
It is our pleasure to thank Nima Arkani-Hamed, Carolina Figueiredo and Fan Zhu for inspiring discussions and collaboration on related projects. The work of SH is supported by the National Natural Science Foundation of China under Grant No. 12225510, 11935013, 12047503, 12247103, and by the New Cornerstone Science Foundation through the XPLORER PRIZE. The work of CS is supported by China Postdoctoral Science Foundation under Grant No. 2022TQ0346, and the National Natural Science Foundation of China under Grant No. 12347146.

\bibliographystyle{apsrev4-1}
\bibliography{fac}

\onecolumn
\clearpage
\begin{appendix}
    \section{Review of factorizations near zeros}
    \label{sec_reviewFac}
    In this appendix, we present a minimal review of the kinematic mesh~\cite{Arkani-Hamed:2019vag}, essential for understanding the factorizations near zeros initially introduced in~\cite{Arkani-Hamed:2023swr}.
    
\begin{figure}[h]
        \centering
        \includegraphics[width=\linewidth]{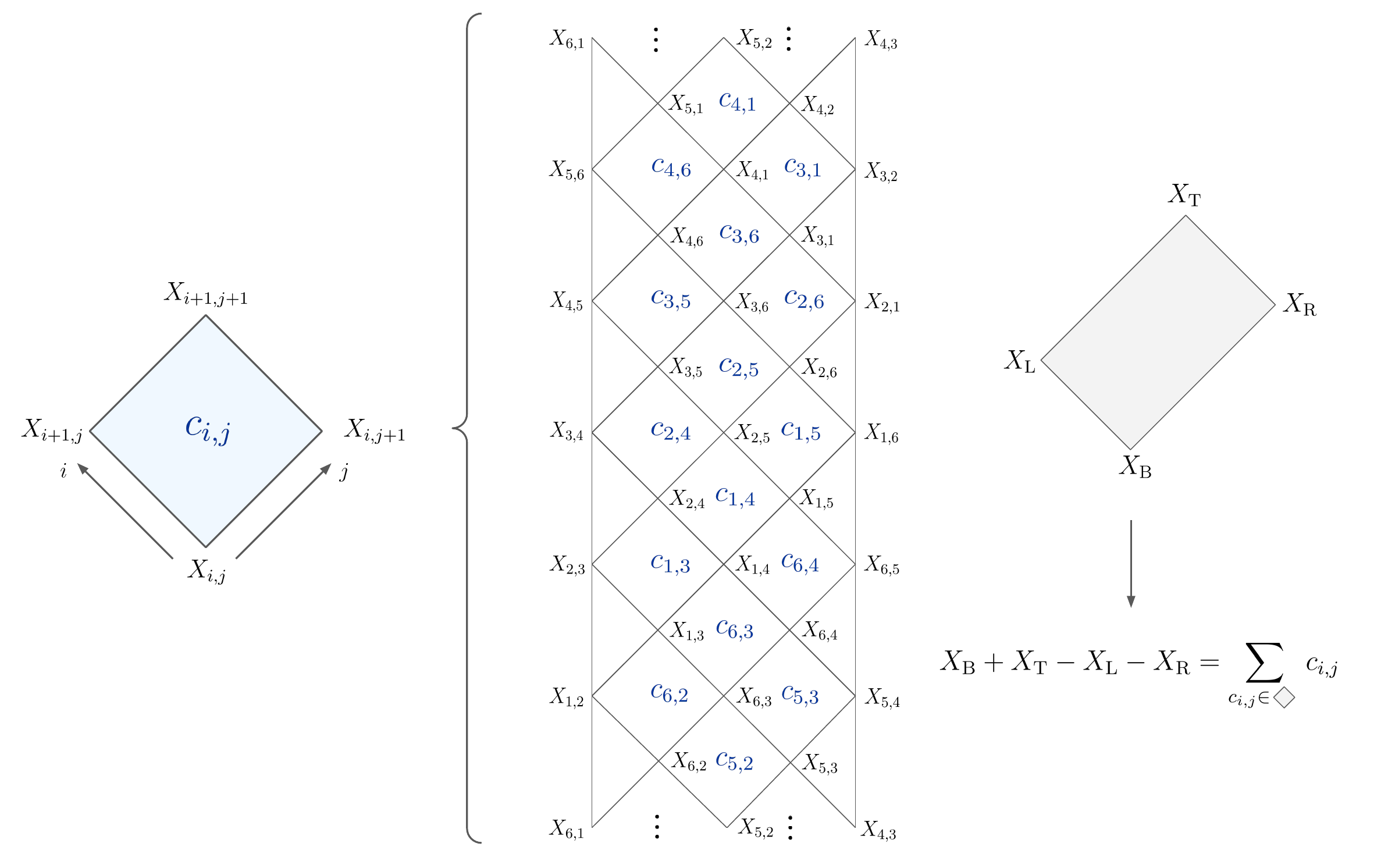}
        \caption{6-point kinematic mesh from~\cite{Arkani-Hamed:2023swr}.}
        \label{fig:6mesh}
    \end{figure}

    A kinematic mesh is associated with a specific ordering of particles: without losing generality, we focus on the canonical ordering $(1,2, \cdots, n)$.
    We define the planar (cyclic) $X_{i,j}$ variables for this ordering: 
    \begin{equation}
        X_{i,j} = (p_i + \dots +p_{j{-}1})^2.
    \end{equation}
    We note that $X_{i,i+1} = p_i^2 = 0$ due to the on-shell conditions, and the remaining $\frac{n(n-1)}{2} - n$ non-vanishing $X_{i,j}$ form of a complete basis of the kinematic space, so that we can express all the other Mandelstam variables in terms of them,
    \begin{equation}
        c_{i,j} := -s_{i,j} = X_{i,j} + X_{i+1,j+1} - X_{i,j+1} - X_{i+1,j},
        \label{eq:ceq}
    \end{equation}
    where $c_{i,j}$ is conventionally used in the kinematic mesh.
    To build up the mesh, one associates a square to each $c_{i,j}$, and the $X_{i,j}$'s in \eqref{eq:ceq} to the vertices of the square~(By custom, it is rotated by $45^\circ$; see figure \ref{fig:6mesh} on the left). 
    Gluing the squares with the same $X_{i,j}$ vertices together, we form a square grid tilted by $45^\circ$. 
    The vertices on boundaries are related to $X_{i,i+1}=0$. 
    In figure \ref{fig:6mesh}, we present the mesh for the 6-point kinematics. 
    Once again, we stress that all the planar variables $X_{i,j}$ are associated with grid points, and the {\it non-planar} dot products of momenta--$c_{i,j}$'s with non-adjacent $i, j$--correspond to the square tiles. 
    The mesh extends infinitely but reflects the cyclic symmetry of the problem by an interesting ``Mobius'' symmetry, where we identify $X_{i,j} = X_{j,i}$ and $c_{i,j}=c_{j,i}$.
    
    Let us proceed to introduce the zeros and factorizations of amplitudes proposed in~\cite{Arkani-Hamed:2023swr}.
    
    \begin{figure}[t]
        \centering
        \includegraphics[width=0.8\textwidth]{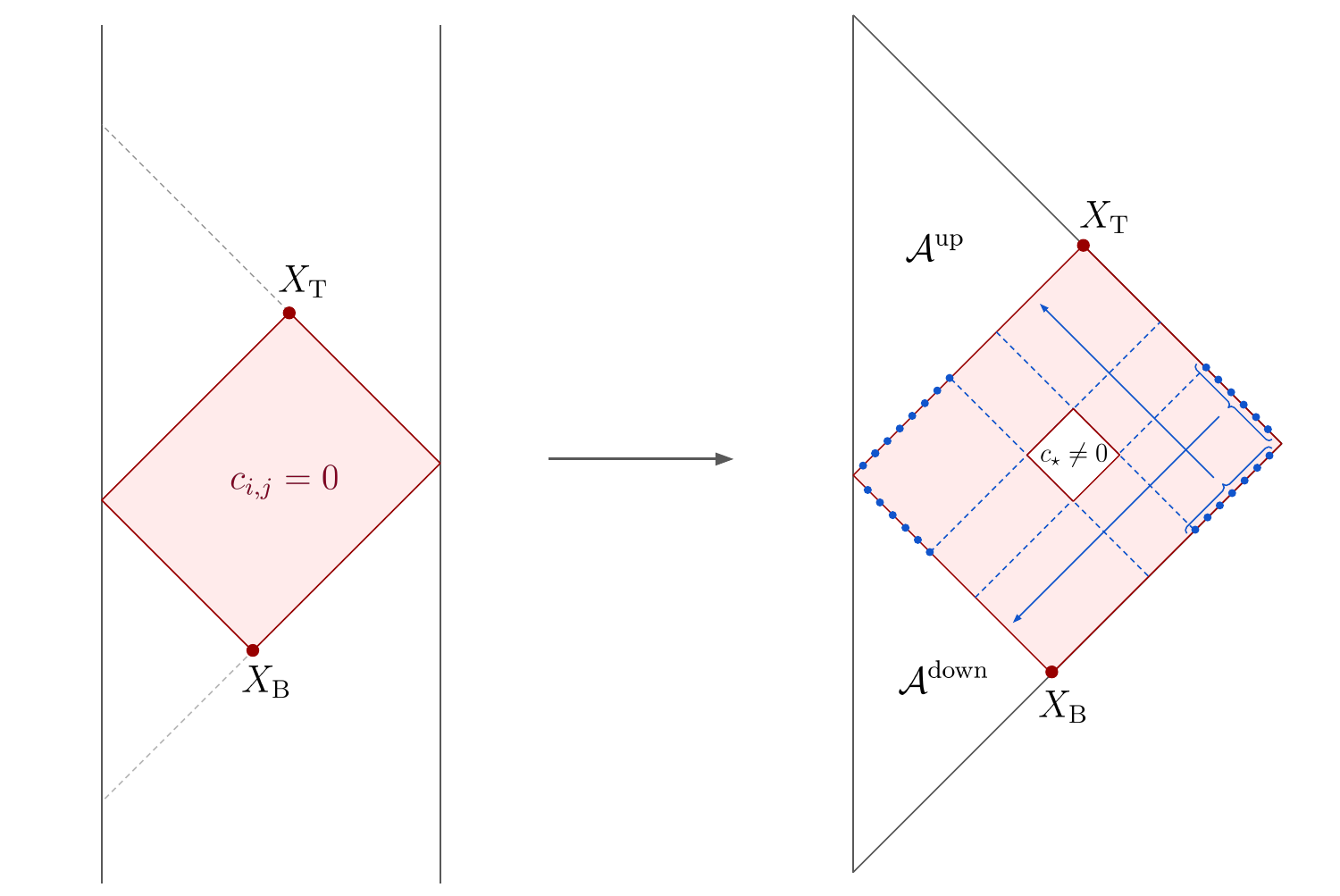}
        \caption{Zeros (left) and Factorizations with associated kinematic shifts (right)~\cite{Arkani-Hamed:2023swr}.}
        \label{fig:ZerosFactGS}
    \end{figure}
    \paragraph{Zeros} For simplicity, we consider an $n$-point tree-level amplitude in Tr$(\phi^3)$ theory. 
    A zero can be found through the following steps:
    (1) Draw the corresponding $n$-point kinematic mesh.
    (2) Pick a point in the mesh, which is associated with a planar variable $X_B$.
    (3) Find the causal diamond anchored in this variable: follow the two light rays starting at $X_B$, let them bounce on the boundaries of the mesh, and meet again at some other point, $X_T$. The region enclosed in the process is called a causal diamond;
    (4) Set all the $c_{i,j}$ inside this causal diamond to zero.
    This will make the amplitude vanish (see Figure~\ref{fig:ZerosFactGS}).
    
    \paragraph{Factorizations} In \cite{Arkani-Hamed:2023swr}, the authors claim that the amplitude factorizes into three pieces of lower-point objects if we turn back on one of the $c$'s inside the zero causal diamond, $c_\star \neq 0$ - hence the term ``factorization near zeros'' (see figure \ref{fig:ZerosFactGS}):
    \begin{equation}
        \mathcal{M}_n(c_\star \neq 0)=\mathcal{M}_4(X_{\text{B}},X_{\text{T}}) \times \mathcal{M}^{\text{down}}\times \mathcal{M}^{\text{up}}.
    \end{equation}
    where $X_B+X_T=c_\star$. For tr$(\phi^3)$, the 4-point amplitude simply reads
    \begin{equation} \label{eq:fac4ptphi3}
        \mathcal{M}_4^{\phi^3}(X_{\text{B}},X_{\text{T}})=\frac{1}{X_{\text{B}}}+\frac{1}{X_{\text{T}}},
    \end{equation}
    and the ``up'' and ``dwon'' amplitudes are also tr$(\phi^3)$ amplitude with specific kinematic shifts (see Figure~\ref{fig_ManMatrix2mesh} in the main text), or equivalently, the tr$(\phi^3)$ current with one off-shell leg. 
    For NLSM and YMS, there are two cases, say ``pure $\times$ pure" $\times$ ${\cal M}^{\phi^3}_4(X_{\text{B}},X_{\text{T}})$, and ``mixed $\times$ mixed" $\times$ ${\cal M}^{\rm NLSM/YMS}_4(X_{\text{B}},X_{\text{T}})$. 
    For the former, the 4-point amplitude is the same as~\eqref{eq:fac4ptphi3}, and the pure NLSM/YMS amplitudes are equally shifted as the tr$(\phi^3)$ cases~\cite{Arkani-Hamed:2023swr}. 
    For the later, the 4-point amplitudes read
    \begin{equation} 
        \mathcal{M}_4^{\rm NLSM}(X_{\text{B}},X_{\text{T}})=X_{\text{B}}+X_{\text{T}}, \qquad  \mathcal{M}_4^{\rm YMS}(X_{\text{B}},X_{\text{T}})=1+\frac{X_{\text{T}}}{X_{\text{B}}}.
    \end{equation}
    The shifted mixed amplitudes for NLSM are given in~\cite{Arkani-Hamed:2023swr}, while for the YMS, there are new ingredients as we will illustrate in~\cite{toappear}.
    
    \section{Examples for $2$-splits and factorizations near zeros}

    In this appendix, we offer concrete examples for 2-splitting and the further factorization of amplitudes. For simplicity, we only consider field theories including bi-adjoint $\phi^3$, NLSM, YM, and GR.
    
    \paragraph{Bi-adjoint $\phi^3$} For starters, let us consider a $6$-point bi-adjoint $\phi^3$ amplitude ${\cal M}^{\phi^3}(\alpha|\beta)$, and choose $\{i,j,k\}=\{1,4,6\}$, $A=\{2,3\}$, and $B=\{5\}$ to construct our 2-split kinematics. 
    In other words, we set $s_{2,5}=s_{3,5}=0$.
    If both orderings $\alpha$ and $\beta$ are canonical, we observe the expected splitting,
    \begin{align}
        \label{eq_6ptphi3}
        {\cal M}^{\phi^3}(1,2,\ldots,6|1,2,\ldots,6)
        \xrightarrow[]{s_{2,5}=s_{3,5}=0}& 
        \left(\frac{1}{s_{1,2} s_{1,2,3}}+\frac{1}{s_{2,3} s_{1,2,3}}+\frac{1}{s_{2,3} s_{2,3,4}}+\frac{1}{s_{3,4} s_{2,3,4}}+\frac{1}{s_{1,2} s_{3,4}} \right) \times 
        \left(\frac{1}{s_{4,5}}+\frac{1}{s_{5,6}} \right)
        \nonumber \\ 
        =&\ {\cal J}^{\phi^3}(1,2,3,4; \kappa | 1,2,3,4; \kappa) \times {\cal J}^{\phi^3}(1,4,5; \kappa^\prime | 1,4,5; \kappa^\prime),
    \end{align}
    where $p_\kappa = -\sum_{\alpha=1}^{4} p_\alpha$, and $p_{\kappa^\prime} = -p_1 -p_4-p_5$ to restore momentum conservation.
    The latter four-point current with an off-shell leg reads
    \begin{equation}
        {\cal J}^{\phi^3} (1,4,5;\kappa^\prime) 
        = \frac{1}{s_{4,5}} + \frac{1}{s_{5,\kappa'}- p_{\kappa^\prime}^2}
        = \frac{1}{s_{4,5}} + \frac{1}{s_{5,6}},
    \end{equation}
    We note that the pole associated with $s_{5,\kappa^\prime}$ is massive, but under our 2-split kinematics it simplifies to $s_{5,6}$~\cite{Naculich:2015zha}.
    The same simplification occurs for the five-point current, leaving no massive pole in \eqref{eq_6ptphi3}.
    To demonstrate the factorization near zeros, we further choose $m=4$, such that $s_{3,6} = 0$.
    The amplitude factorizes as
    \begin{equation}
        \begin{aligned}
            \eqref{eq_6ptphi3}
            \xrightarrow[]{s_{3,6}=0}& 
            \left(\frac{1}{s_{3,4}}+\frac{1}{s_{2,3}} \right) \times \left(\frac{1}{s_{2,3,4}}+\frac{1}{s_{1,2,3}} \right) \times 
            \left(\frac{1}{s_{4,5}}+\frac{1}{s_{5,6}} \right) \\
            =&\ {\cal J}^{\phi^3} (3,4,1;\rho | 3,4,1;\rho) 
            \times {\cal J}^{\phi^3} (1,\rho^\prime,4,\kappa | 1,\rho^\prime,4,\kappa) 
            \times {\cal J}^{\phi^3} (1,4,5;\kappa^\prime | 1,4,5;\kappa^\prime),
        \end{aligned}
    \end{equation}
    where $p_\rho = -p_1-p_3-p_4$, $p_{\rho^\prime} = -p_1-p_4-p_{\kappa}$, and ${\cal J}^{\phi^3} (1,\rho^\prime,4,\kappa | 1,\rho^\prime,4,\kappa)$ is the universal $4$-point object,
    For non-canonical color orderings compatible with the split kinematics, the amplitude splits and factorizes in the same way.
    For instance, if we swap the position of $2,3$ in the second ordering, ${\cal M}^{\phi^3}(\alpha|\beta)$ becomes
    \begin{multline}
        {\cal M}^{\phi^3}(1,2,3,4,5,6|1,3,2,4,5,6)
        \xrightarrow[]{s_{2,5}=s_{3,5}=0}
        \ {\cal J}^{\phi^3} (1,2,3,4;\kappa | 1,3,2,4;\kappa) 
        \times {\cal J}^{\phi^3}(1,4,5;\kappa^\prime|1,4,5;\kappa^\prime) \\
        \xrightarrow[]{s_{3,6}=0}
        {\cal J}^{\phi^3}(3,4,1;\rho|4,1,3;\rho) \times 
        {\cal J}^{\phi^3}(1,\rho^\prime,4,\kappa|1,\rho^\prime,4,\kappa) \times 
        {\cal J}^{\phi^3}(1,4,5;\kappa^\prime|1,4,5;\kappa^\prime).
    \end{multline}
    More non-trivial examples at 10 points are 
    \begin{gather}
        \begin{aligned}
            {\cal M}^{\phi^3} (1,2,\ldots,10 &| 1,2,\ldots,10)
            \xrightarrow[]{i=2,j=7,k=10} 
            \mathcal{J}^{\phi^3} (2,\ldots,7; \kappa | 2,\ldots,7; \kappa)
            \mathcal{J}^{\phi^3} (1,2,7,8,9; \kappa^\prime | 1,2,7,8,9; \kappa^\prime) \\ 
            &\xrightarrow[]{m=4} 
            \mathcal{J}^{\phi^3} (5,6,7,2,3;\rho | 5,6,7,2,3;\rho)
            \,\mathcal{J}^{\phi^3} (2,\rho^\prime,7; \kappa | 2,\rho^\prime,7; \kappa)
            \,\mathcal{J}^{\phi^3} (1,2,7,8,9; \kappa^\prime | 1,2,7,8,9; \kappa^\prime) 
        \end{aligned}
        \\
        \begin{aligned}
            {\cal M}^{\phi^3} (1,2,\ldots,10 &| 1,\textbf{2},6,5,4,3,\textbf{7}, 10,9,8)
            \xrightarrow[]{i=2,j=7,k=10} 
            \mathcal{J}^{\phi^3} (2,\ldots,7; \kappa | \textbf{2},6,5,4,3,\textbf{7}; \kappa)
            \mathcal{J}^{\phi^3} (1,2,7,8,9; \kappa^\prime | 9,8,1,\textbf{2},\textbf{7}, \kappa^\prime), \\ 
            &\xrightarrow[]{m=4}
            \mathcal{J}^{\phi^3} (5,6,7,2,3;\rho | 3,7,2,6,5;\rho)
            \,\mathcal{J}^{\phi^3} (2,\rho^\prime,7; \kappa | 2,\rho^\prime,7; \kappa)
            \,\mathcal{J}^{\phi^3} (1,2,7,8,9; \kappa^\prime | 9,8,1,\textbf{2},\textbf{7}, \kappa^\prime),
        \end{aligned}
    \end{gather}
    
    \paragraph{NLSM} Analogously, imposing the 2-split kinematic conditions to the $6$-point NLSM amplitude yields 
    \begin{equation}
        \begin{aligned}
            {\cal M}^\mathrm{NLSM}(1,2,\ldots,6)
            \xrightarrow[]{s_{2,5}=s_{3,5}=0}& \left(1-\frac{s_{1,2}}{s_{1,2,3}}-\frac{s_{2,3}}{s_{1,2,3}}-\frac{s_{2,3}}{s_{2,3,4}}-\frac{s_{3,4}}{s_{2,3,4}} \right) \times s_{1,5}\\
            =&\  {\cal J}^{\mathrm{NLSM}+\phi^3}(1^\phi,2,3,4^\phi; \kappa^{\phi})\;\times {\cal J}^\mathrm{NLSM}(1,4,5;\kappa^\prime) \\
            \xrightarrow[]{s_{3,6}=0}&\ s_{1,3} \times \left( \frac{1}{s_{1,2,3}} +\frac{1}{s_{2,3,4}} \right) \times s_{1,5}\\  
            =&\  {\cal J}^{\mathrm{NLSM}}(3,4,1;\rho) \times {\cal J}^{\phi^3}(1,\rho^\prime,4,\kappa) \times {\cal J}^\mathrm{NLSM}(1,4,5;\kappa^\prime).
        \end{aligned}
    \end{equation}
    A less trivial case at $10$ points allows both even-even and odd-odd factorization,
    \begin{gather}
        \begin{aligned}
            {\cal M}^\mathrm{NLSM}(1,2,\ldots,10)& \xrightarrow[]{i=2,j=7,k=10} 
            {\cal J}^{\mathrm{NLSM}+\phi^3}(2^{\phi},3,4,5,6,7^{\phi}; \kappa^\phi)\; 
            {\cal J}^{\mathrm{NLSM}}(1,2,7,8,9; \kappa^{\prime})\\
            & \xrightarrow[]{m=4} {\cal J}^{\mathrm{NLSM}}(5,6,7,2,3; \rho) 
            {\cal J}^\mathrm{\phi^3}(2,\rho^\prime,7,\kappa)\; 
            {\cal J}^{\mathrm{NLSM}}(1,2,5,8,9; \kappa^{\prime}).
        \end{aligned}
        \\
        \begin{aligned}
            {\cal M}^\mathrm{NLSM}(1,2,\ldots,10)& \xrightarrow[]{i=2,j=6,k=10} {\cal J}^\mathrm{NLSM}(2,3,4,5,6; \kappa)\; {\cal J}^{\mathrm{NLSM}+\phi^3}(1,2^\phi,6^\phi,7,8,9; \kappa^{\prime\, \phi})\\
            & \xrightarrow[]{m=4} {\cal J}^{\mathrm{NLSM}+\phi^3}(5,6^\phi,2^\phi,3; \rho^\phi) {\cal J}^\mathrm{NLSM}(2,\rho^\prime,6,\kappa)\; {\cal J}^{\mathrm{NLSM}+\phi^3}(1,2^\phi,6^\phi,7,8,9; \kappa^{\prime\, \phi}),
        \end{aligned}
    \end{gather}
    where the second (odd-odd) factorization gives rise to a four-point NLSM object, ${\cal J}^\mathrm{NLSM} (2,\rho^\prime,6,\kappa) = -s_{2,\kappa} {-} s_{6,\kappa} = {-}s_{3,4,5,6} {-}s_{2,3,4,5}$.
    
    \paragraph{Yang-Mills} As discussed in the main text, for YM, in addition to the usual split condition, we also need to impose appropriate conditions involving the polarizations.
    For instance, at $7$ points, we pick $\{i=1, j=4, k=7\}$, and set
    \begin{equation}
        \label{eq_7ptYMkin}
        s_{a,b}
        = \epsilon_{a} \cdot \epsilon_{b^\prime} 
        = p_{a} \cdot \epsilon_{b^\prime} 
        = \epsilon_{a} \cdot p_{b} 
        = 0, \quad
        \text{ for } \quad
        a \in \{2,3\},\  
        b \in \{5,6\},\
        b^\prime \in \{5,6\}\cup \{1,4,7\},
    \end{equation}
    such that the YM amplitude becomes
    \begin{equation}
        \label{eq_7ptYM2split}
        \begin{aligned}
            {\cal M}^\mathrm{YM}(1,2,3,4,5,6,7)& \xrightarrow[]{i=1,j=4,k=7} 
            {\cal J}^{\mathrm{YM}+\phi^3} (1^\phi,2,3,4^\phi; \kappa^\phi)\; 
            {\cal J}^{\mathrm{YM}} (1,4,5,6; \kappa^{\prime})_{\mu} \epsilon_{7}^{\mu} ,
        \end{aligned}
    \end{equation}
    where $\epsilon_7^\mu$ should be reinterpreted as associated with the off-shell momentum $\kappa^\prime$. 
    Furthermore, by further splitting either the mixed or the pure current, we can arrive at different 3-factorizations.
    For the former case, we choose $m=2$ and impose $\epsilon_{3}\cdot \epsilon_{2} = p_{3} \cdot \epsilon_{2} = \epsilon_{3}\cdot p_{7} = p_{3}\cdot p_{7}=0$, such that
    \begin{equation}
        \eqref{eq_7ptYM2split} \xrightarrow[]{m=2}   
        {\cal J}^{\mathrm{YM}+\phi^3} (1^\phi,3,4^\phi; \rho^\phi)
        \times {\cal J}^{\mathrm{YM}+\phi^3} (1^\phi,\rho^{\prime},4^\phi, \kappa^\phi)_{\nu} \epsilon^{\nu}_{2}\,
        \times {\cal J}^{\mathrm{YM}} (1,4,5,6; \kappa^{\prime})_{\mu} \epsilon_{7}^{\mu} .
    \end{equation}
    For the latter, with $m=5$ and setting $p_6 \cdot p_{7} = \epsilon_{6} \cdot \epsilon_{b^\prime}= p_6 \cdot \epsilon_{b^\prime} = \epsilon_6\cdot p_{7} =0$ for $b^\prime \in \{1,4,5,7\}$, we have
    \begin{equation}
        \eqref{eq_7ptYM2split} \xrightarrow[]{m=5}   
        {\cal J}^{\mathrm{YM}+\phi^3} (1^\phi,2,3,4^\phi; \kappa^\phi)\; 
        \times {\cal J}^{\mathrm{YM}} (1,4, \rho^{\prime},\kappa)_{\nu\mu} \epsilon^{\nu}_{5} \epsilon_{7}^{\mu} \,
        \times {\cal J}^{\mathrm{YM}} (6, 1^\phi,4^\phi; \rho^{\phi}).
    \end{equation}
    
    Surprisingly, for YM, an even simpler split kinematic with set $B$ being empty is possible.
    This is non-trivial since one still needs to decouple the polarizations of $\{i,j,k\}$ from the particles in set $A$ to observe the 2-split behavior.
    It becomes evident even at $4$ points, where, with $\epsilon_2 \cdot \epsilon_{b^\prime}=p_2 \cdot \epsilon_{b^\prime}=0$ for $b^\prime \in \{1,3,4\}$ the YM amplitude splits as
    \begin{equation}
        \begin{aligned}
            {\cal M}^\mathrm{YM}(1,2,3,4) \xrightarrow[]{i=1,j=3,k=4}
            &\left(-\frac{p_3\cdot \epsilon _2}{s_{2,3}}
            +\frac{p_1\cdot \epsilon _2}{s_{1,2}}\right)
            \left(
            \epsilon_1 \!\cdot\! \epsilon _3\, p_{3}^{\mu}
            + \epsilon_3 \!\cdot\! p_1 \, \epsilon_{1}^{\mu}
            + \epsilon_1 \!\cdot\! p_{\kappa^{\prime}} \, \epsilon_{3}^{\mu} \right)
            \epsilon_{4,\mu} \\
            &=
            {\cal J}^{\mathrm{YM}+\phi^3}(1^\phi,2,3^\phi; \kappa^\phi)\; 
            {\cal J}^{\mathrm{YM}}(1,3; \kappa^{\prime})_{\mu} \epsilon_{4}^{\mu}.
        \end{aligned}
    \end{equation}
    
    \paragraph{Gravity} For graviton amplitudes, we can choose to impose the constraints on the two polarization vectors $\epsilon_\mu, \tilde{\epsilon}_\mu$ of $\{i,j,k\}$ independently.
    Let us take a $7$-point amplitude as an example.
    If we assign both polarizations to the same side, \textit{i.e.}, \eqref{eq_7ptYMkin} applies identically to $\epsilon_\mu, \tilde{\epsilon}_\mu$, the amplitude splits in a similar way as the YM one~\eqref{eq_7ptYM2split},
    \begin{equation}
        \label{eq_6ptGR2split}
        \begin{aligned}
            {\cal M}^\mathrm{GR}(1,2,3,4,5,6,7)& \xrightarrow[]{i=1,j=4,k=7} 
            {\cal J}^{\mathrm{GR}+\phi^3} (1^\phi,2,3,4^\phi; \kappa^\phi)\; 
            {\cal J}^{\mathrm{GR}} (1,4,5,6; \kappa^{\prime})_{\mu\nu} \epsilon_{7}^{\mu} \tilde{\epsilon}_{7}^{\nu} ,
        \end{aligned}
    \end{equation}
    where we note that the second term is pure GR.
    Alternatively, if we adopt \eqref{eq_7ptYMkin} only for $\epsilon_\mu$, and enforce the following conditions for $\tilde{\epsilon}_\mu$,
    \begin{equation}
        \label{eq_6ptGRkin}
        \tilde{\epsilon}_{b} \cdot \tilde{\epsilon}_{a^\prime}
        = p_{a} \cdot \tilde{\epsilon}_{b} 
        =  p_{b} \cdot \tilde{\epsilon}_{a^\prime}
        = 0, \quad
        \text{ for } \quad
        a \in \{2,3\},\  
        b \in \{5,6\},\
        a^\prime \in \{2,3\}\cup \{1,4,7\},
    \end{equation}
    then we obtain two mixed currents, each with three gluons and the remaining particles being gravitons
    \begin{equation}
        \label{eq_6ptGR2split2}
        \begin{aligned}
            {\cal M}^\mathrm{GR}(1,2,3,4,5,6,7)& \xrightarrow[]{i=1,j=4,k=7} 
            \epsilon_{7}^{\mu} {\cal J}^{\mathrm{EYM}} (1^g,2,3,4^g; \kappa^g)_{\mu} \; 
            {\cal J}^{\mathrm{EYM}} (1^g,4^g,5,6; \kappa^{\prime g})_{\nu} \tilde{\epsilon}_{7}^{\nu}.
        \end{aligned}
    \end{equation}

    \section{Splitting bosonic string correlators}
    In this appendix, we provide a brief derivation of the split of the bosonic string correlator.
    The gauge-fixed bosonic string correlators for $n$-gluon scattering are given by:
    \begin{equation}
        \mathcal{C}_n(\epsilon,p,z)=\Delta_{i,j,n} \sum_{r=0}^{\lfloor n/2 \rfloor{+}1} \sum_{\{g,h\}, \{l\}} \prod_{s}^r W_{g_s, h_s} \prod_{t}^{n-2r} V_{l_t}, \quad V_{i}:= \sum_{j\neq i}^n \frac{\epsilon_i \cdot p_j}{z_{i,j}}, \quad W_{i,j}:=\frac{\epsilon_i \cdot \epsilon_j}{ z_{i,j}^2},
    \end{equation}
    where we have a summation over all partitions of $\{1,2,\cdots, n\}$ into $r$ pairs $\{g_s, h_s\}$ and $n-2r$ singlets $l_t$, each summand given by the product of $W$'s and $V$'s. For example, the $n=3$ case reads $\Delta_{i,j,3}(V_1 V_2 V_3+ W_{1,2} V_3 + W_{2,3} V_1+ W_{1,3} V_2)$, and for $n=4$ we have
    \begin{equation}
        \mathcal{C}_4(\epsilon,p,z)=\Delta_{i,j,4}\left(V_1 V_2 V_3 V_4+ \left(W_{1,2} V_3 V_4+ {\rm perm.} \right)+ \left( W_{1,2} W_{3,4} + {\rm perm.}\right)\right),
    \end{equation}
    with the permutation exhausts all $6$ terms of the form $W V V$ and $3$ terms of the form $W W$. Now let us impose the splitting conditions~\eqref{eq_pol}, which enforce
    \begin{equation}
        W_{a,b'}=0,\quad  V_{a}= \sum_{c \neq a, c\notin B} \frac{\epsilon_{a} \cdot p_c}{z_{a,c}},\quad V_{b'}= \sum_{c \neq b', c\notin A} \frac{\epsilon_{b'} \cdot p_c}{z_{b',c}}
    \end{equation}
    Therefore, the polarization of $A$ and $B'$ completely decouple, and the summations in $V_a, V_{b'}$ only involve $A\cup \{i,j,\kappa\}$ or $B \cup \{i,j,\kappa'\}$ (with $\kappa,\kappa'$ missing since we have fixed $z_{\kappa},z_{\kappa'}\to \infty$), respectively.
    As a consequence, the bosonic string correlator behaves as
    \begin{equation}
        \begin{aligned}
            \mathcal{C}_n(\epsilon,p,z) &\to \Delta_{i,j,\kappa}\mathrm{PT}(i,j,\kappa)\left(\sum_{r=0}^{\lfloor |A|/2 \rfloor{+}1} \sum_{\{g,h\}, \{l\} \in A} \prod_{s}^r W_{g_s, h_s} \prod_{t}^{|A|-2r} V_{l_t}\right) \\
            & \times
            \Delta_{i,j,\kappa'}\left(\sum_{r=0}^{\lfloor |B'|/2 \rfloor{+}1} \sum_{\{g,h\}, \{l\} \in B'} \prod_{s}^r W_{g_s, h_s} \prod_{t}^{|B'|-2r} V_{l_t}\right),
        \end{aligned}
    \end{equation}
    which results in the splitting of a bosonic string amplitude: the first line corresponds to a mixed amplitude with $|A|$ gluons and $\{i,j, \kappa\}$ being three $\phi's$~\cite{He:2019drm}; the second line represents a pure gluons amplitude with external legs in $B \cup \{i,j,\kappa'\}$.

\end{appendix}

\end{document}